\documentclass[twoside]{article}
\usepackage{myespcrc2}

\usepackage{graphicx}
\usepackage[figuresright]{rotating}

\title{Results from the Antarctic Muon and Neutrino Detector Array}

\author{D.F. Cowen for the AMANDA Collaboration (invited talk given at Neutrino 2002, Munich, Germany): \\
J.~Ahrens\address[MAINZ]{Institute of Physics, University of Mainz, Staudinger Weg 7, D-55099 Mainz, Germany},
X.~Bai\address[BARTOL]{Bartol Research Institute, University of Delaware, Newark, DE 19716, USA}, 
S.W.~Barwick\address[IRVINE]{Dept. of Physics and Astronomy, University of California, Irvine, CA 92697, USA}, 
T.~Becka\addressmark[MAINZ], 
K.-H.~Becker\address[WUPPERTAL]{Fachbereich 8 Physik, BUGH Wuppertal, D-42097 Wuppertal, Germany},
E.~Bernardini\address[DESY]{DESY-Zeuthen, D-15735 Zeuthen, Germany},
D.~Bertrand\address[ULB]{Universit\'e Libre de Bruxelles, Science Faculty CP230, B-1050 Brussels, Belgium}, 
F.~Binon\addressmark[ULB], 
A.~Biron\addressmark[DESY],
S.~B\"oser\addressmark[DESY], 
O.~Botner\address[UPPSALA]{Division of High Energy Physics, Uppsala University, S-75121 Uppsala, Sweden},
O.~Bouhali\addressmark[ULB], 
T.~Burgess\address[STOCKHOLM]{Dept. of Physics, Stockholm University, SCFAB, SE-10691 Stockholm, Sweden},
S.~Carius\address[KALMAR]{Dept. of Technology, Kalmar University, S-39182 Kalmar, Sweden},
T.~Castermans\address[MONS]{University of Mons-Hainaut, 7000 Mons, Belgium},
D.~Chirkin\address[BERKELEY]{Dept. of Physics, University of California, Berkeley, CA 94720, USA},
J.~Conrad\addressmark[UPPSALA], 
J.~Cooley\address[MADISON]{Dept. of Physics, University of Wisconsin, Madison, WI 53706, USA},
D.F.~Cowen\address[PSU]{Dept. of Physics, Pennsylvania State University, University Park, PA 16802, USA},
A.~Davour\addressmark[UPPSALA], 
C.~De~Clercq\address[VUB]{Vrije Universiteit Brussel, Dienst ELEM, B-1050 Brussels, Belgium},
T.~DeYoung\addressmark[MADISON]\thanks{Present address: Santa Cruz Institute for Particle Physics, University of California, Santa Cruz, CA 95064, USA.}, 
P.~Desiati\addressmark[MADISON], 
J.-P.~Dewulf\addressmark[ULB], 
P.~Doksus\addressmark[MADISON], 
P.~Ekstr\"om\addressmark[STOCKHOLM],
T.~Feser\addressmark[MAINZ], 
T.K.~Gaisser\addressmark[BARTOL], 
R.~Ganupati\addressmark[MADISON],
M.~Gaug\addressmark[DESY], 
H.~Geenen\addressmark[WUPPERTAL],
L.~Gerhardt\addressmark[IRVINE],
A.~Goldschmidt\address[LBNL]{Lawrence Berkeley National Laboratory, Berkeley, CA 94720, USA},
A.~Hallgren\addressmark[UPPSALA], 
F.~Halzen\addressmark[MADISON], 
K.~Hanson\addressmark[MADISON], 
R.~Hardtke\addressmark[MADISON], 
T.~Hauschildt\addressmark[DESY], 
M.~Hellwig\addressmark[MAINZ], 
P.~Herquet\addressmark[MONS],
G.C.~Hill\addressmark[MADISON], 
P.O.~Hulth\addressmark[STOCKHOLM],
K.~Hultqvist\addressmark[STOCKHOLM],
S.~Hundertmark\addressmark[STOCKHOLM], 
J.~Jacobsen\addressmark[LBNL], 
A.~Karle\addressmark[MADISON], 
L.~K\"opke\addressmark[MAINZ], 
K.~Kuehn\addressmark[IRVINE],
M.~Kowalski\addressmark[DESY], 
J.I.~Lamoureux\addressmark[LBNL], 
H.~Leich\addressmark[DESY], 
M.~Leuthold\addressmark[DESY], 
P.~Lindahl\addressmark[KALMAR], 
I.~Liubarsky\addressmark[MADISON], 
J.~Madsen\address[UWRF]{Physics Dept., University of Wisconsin, River Falls, WI 54022, USA},
K.~Mandli\addressmark[MADISON], 
P.~Marciniewski\addressmark[UPPSALA], 
H.S.~Matis\addressmark[LBNL], 
C.P.~McParland\addressmark[LBNL], 
T.~Messarius\addressmark[WUPPERTAL],
Y.~Minaeva\addressmark[STOCKHOLM], 
P.~Mio\v{c}inovi\'c\addressmark[BERKELEY], 
R.~Morse\addressmark[MADISON], 
R.~Nahnhauer\addressmark[DESY],
T.~Neunh\"offer\addressmark[MAINZ], 
P.~Niessen\addressmark[VUB], 
D.R.~Nygren\addressmark[LBNL], 
H.~Ogelman\addressmark[MADISON], 
Ph.~Olbrechts\addressmark[VUB], 
C.~P\'erez~de~los~Heros\addressmark[UPPSALA], 
A.C.~Pohl\addressmark[KALMAR], 
P.B.~Price\addressmark[BERKELEY], 
G.T.~Przybylski\addressmark[LBNL], 
K.~Rawlins\addressmark[MADISON], 
E.~Resconi\addressmark[DESY], 
W.~Rhode\addressmark[WUPPERTAL], 
M.~Ribordy\addressmark[DESY], 
S.~Richter\addressmark[MADISON], 
J.~Rodr\'\i guez~Martino\addressmark[STOCKHOLM], 
D.~Ross\addressmark[IRVINE], 
H.-G.~Sander\addressmark[MAINZ], 
T.~Schmidt\addressmark[DESY], 
D.~Schneider\addressmark[MADISON],
K.~Schinarakis\addressmark[WUPPERTAL], 
R.~Schwarz\addressmark[MADISON], 
A.~Silvestri\addressmark[IRVINE], 
M.~Solarz\addressmark[BERKELEY], 
G.M.~Spiczak\addressmark[UWRF], 
C.~Spiering\addressmark[DESY], 
D.~Steele\addressmark[MADISON], 
P.~Steffen\addressmark[DESY], 
R.G.~Stokstad\addressmark[LBNL], 
P.~Sudhoff\addressmark[DESY],
K.-H.~Sulanke\addressmark[DESY], 
I.~Taboada\address{Departamento de F\'{\i}sica, Universidad Sim\'on Bol\'{\i}var, Apdo. Postal 89000, Caracas, Venezuela}, 
L.~Thollander\addressmark[STOCKHOLM], 
S.~Tilav\addressmark[BARTOL], 
W.~Wagner\addressmark[WUPPERTAL],
C.~Walck\addressmark[STOCKHOLM], 
C.~Weinheimer\addressmark[MAINZ], 
C.H.~Wiebusch\addressmark[DESY]\thanks{Present address: CERN, CH-1211, Gen\`eve 23, Switzerland.},
C.~Wiedemann\addressmark[STOCKHOLM],
R.~Wischnewski\addressmark[DESY], 
H.~Wissing\addressmark[DESY], 
K.~Woschnagg\addressmark[BERKELEY], 
G.~Yodh\addressmark[IRVINE], 
S.~Young\addressmark[IRVINE]
}

\begin{document}

\begin{abstract}

\noindent We show new results from both the
older and newer incarnations of AMANDA (AMANDA-B10 and AMANDA-II,
respectively).  These results demonstrate that AMANDA is a functioning,
multipurpose detector with significant physics and astrophysics reach.
They include a new higher-statistics measurement of the atmospheric
muon neutrino flux and preliminary results from searches for a variety
of sources of ultrahigh energy neutrinos: generic point sources,
gamma-ray bursters and diffuse sources producing muons in the
detector, and diffuse sources producing electromagnetic or hadronic
showers in or near the detector.

\vspace{1pc}
\end{abstract}

\maketitle

\section{INTRODUCTION}
 
Ultrahigh energy (UHE) neutrinos with energies in the TeV range and
higher may be produced by a variety of sources.  Particle physics
exotica like WIMPs and topological defects are expected to produce
neutrinos in their annihilation or decay~\cite{uhe-nu-hep-theories},
and models of astrophysical phenomena such as gamma-ray bursts, active
galactic nuclei, supernovae and
microquasars~\cite{uhe-nu-astro-theories} also predict UHE neutrino
fluxes.

AMANDA is sensitive to UHE neutrinos produced by these sources and can
provide some of the most stringent tests to date of UHE neutrino
production models.  More generally, AMANDA and other similar neutrino
telescopes~\cite{other-experiments} open a heretofore unexplored
window on the universe in a region of the energy spectrum bounded
between roughly $10^{12}$~eV and $10^{20}$~eV.  In the somewhat
narrower energy range between roughly $10^{14}$~eV and $10^{19}$~eV,
photons are absorbed by intervening matter and starlight, and
cosmic-ray protons are insufficiently energetic to reach us without
experiencing unknown amounts of curvature in intervening magnetic
fields, leaving neutrinos as the only known particles that can serve
as astronomical messengers.  Neutrino telescopes are also sensitive to
supernova neutrino bursts at neutrino energies of roughly $10^7$~eV.

\section{THE AMANDA DETECTOR}

The AMANDA-B10 high energy neutrino detector consists of 302 optical
modules (OMs) on 10 strings.  Each OM comprises a photomultiplier
tube (PMT) with passive electronics housed in a glass pressure vessel.
The OMs are deployed within a cylindrical volume about 120~m in diameter
and 500~m in height at depths between roughly 1500 and 2000~m below
the surface of the South Pole ice cap.  In this region the optical
properties of the ice are well suited for reconstructing the Cherenkov
light pattern emitted by relativistic charged
particles~\cite{ice-properties}. This light is used to reconstruct
individual events.  An electrical cable provides high voltage to the
PMTs and transmits their signal pulses to the surface electronics.  A
light diffuser ball connected via fiber optic cable to a laser on the
surface is used for calibration purposes.  Copious down-going cosmic
ray muons are also used for calibration purposes.

\begin{figure}[htb]
\vspace{9pt}
% \framebox[55mm]{\rule[-21mm]{0mm}{43mm}}
\includegraphics[scale=0.4]{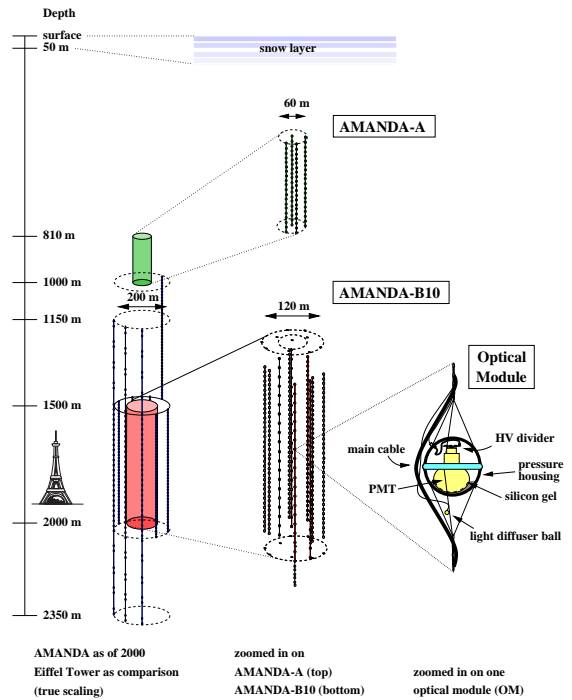}
\caption{AMANDA-B10 consists of 302 optical modules in a cylindrical volume 
120~m diameter and 500~m in height.  To build AMANDA-II, optical
modules were added to bring the count up to 667 OMs in a cylindrical
volume 200~m in diameter.}
\label{fig:amanda-detector}
\end{figure}

In January 2000, AMANDA-B10 was enlarged to a total of 19 strings with
667 OMs to form AMANDA-II.  This new detector is 200~m in diameter and
approximately the same height and depth as AMANDA-B10.
Figure~\ref{fig:amanda-detector} shows a schematic diagram of AMANDA.
Design and construction has begun on IceCube, a kilometer-scale device
with 4800 OMs on 80 strings~\cite{albrechts-contribution}.

\section{ATMOSPHERIC NEUTRINOS AND OTHER PHYSICS WITH AMANDA-B10}

\begin{figure}[htb]
\vspace{9pt}
% \framebox[55mm]{\rule[-21mm]{0mm}{43mm}}
\includegraphics[scale=0.55]{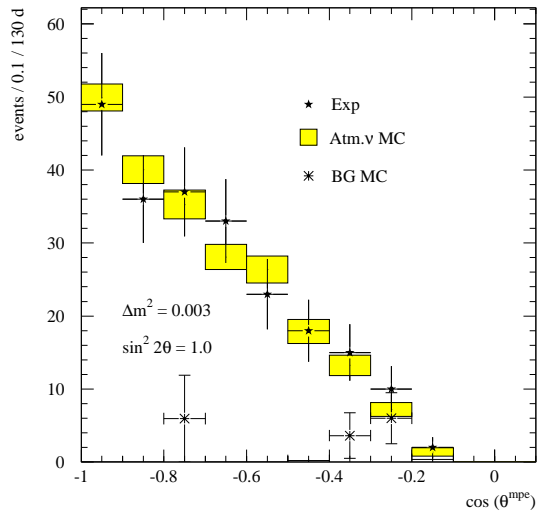}
\caption{Number of upward-going muon events in AMANDA-B10 data from 
         the year 1997, as a function of zenith angle ($\cos{\theta} =
         -1.0$ is vertically up in the detector).  The data are shown
         as dots and the Monte Carlo as boxes.  The simulation was performed
         with the neutrino oscillation parameters as indicated.
         The predicted signal efficiency is
         roughly 4\% and the background level is roughly 10\%, with both
         numbers improving near the vertical and degrading near the horizon.
         Simulations indicate that 90\% of these events lie in an energy range
         given roughly by 66~GeV $< E_\nu <$ 3.4~TeV.}
\label{fig:b10-atmnu}
\end{figure}

AMANDA was shown to be a functioning neutrino telescope by virtue of
its ability to reconstruct upward-going muons induced by atmospheric
muon neutrinos~\cite{nature,b10-atmnu}.  A fraction of the atmospheric
muon neutrinos produced in the northern hemisphere travel through the
earth, interact with the underlying earth or the ice near AMANDA, and
produce a muon which can be detected and reconstructed.  Using data
collected by AMANDA-B10 in 1997, we reconstructed roughly 300
upward-going muons which, as shown in fig.~\ref{fig:b10-atmnu},
are in agreement with the predicted angular distribution.

AMANDA-B10 data has also been used to set competitive limits on
WIMPs~\cite{b10-wimps}, monopoles~\cite{b10-monopoles}, extremely
energetic neutrinos~\cite{b10-ehe}, UHE $\nu_\mu$ point
sources~\cite{b10-point-source} and diffuse fluxes~\cite{b10-diffuse}.

The detector is also sensitive to
bursts of low energy neutrinos from
supernovae~\cite{b10-supernova}.

\section{ATMOSPHERIC NEUTRINOS WITH AMANDA-II}

\begin{figure}[htb]
\vspace{9pt}
% \framebox[55mm]{\rule[-21mm]{0mm}{43mm}}
\includegraphics[scale=0.43,bb=30 30 560 560]{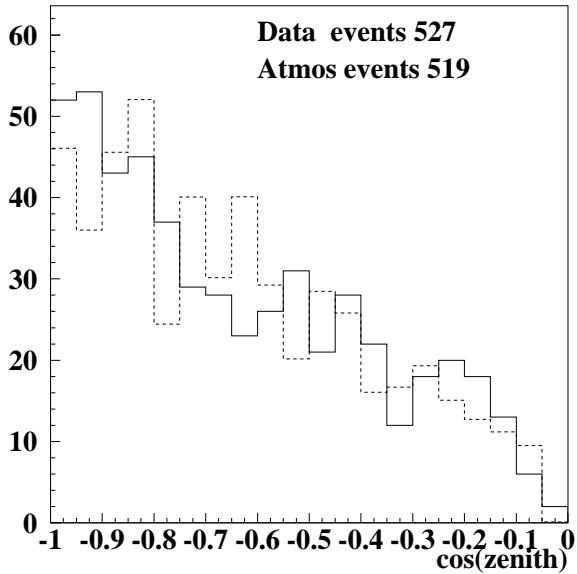}
\caption{Number of upward-going muon events in AMANDA-II data from the 
         year 2000 as a function of zenith angle, using a preliminary
         set of selection criteria.  There are a total of 527 events
         in the data (solid line), and 519 events predicted by the
         atmospheric neutrino Monte Carlo (dashed line).  Simulations
         indicate that these events have an energy range given roughly
         by 100~GeV $< E_\nu <$ 1~TeV.  With more sophisticated selection criteria
         we expect improved response near the horizon.}
\label{fig:aii-atmnu}
\end{figure}

A preliminary analysis of atmospheric neutrino data taken with
AMANDA-II demonstrates the substantially increased power of the
enlarged detector.  Compared to the analysis using AMANDA-B10 data,
fewer selection criteria are required to extract a larger and
qualitatively cleaner sample of atmospheric neutrino-induced muons.
Figure~\ref{fig:aii-atmnu} shows the excellent shape agreement between
data and simulation achieved with a preliminary set of selection
criteria applied.  With more sophisticated selection criteria we
expect to see roughly twice the number of events shown in the figure
(corresponding to 2-3 times more events in AMANDA-II relative to
AMANDA-B10 for equivalent live-times) and we also anticipate improved
angular response close to the horizon.
%
% Live time was 226 days, vs. 130 for B10.  
% Expected 226/130 * 204 = 355, saw 527: 527/355 =1.5, a 50% improvement.
%
%There are 527 events in the sample, more than twice the
%number extracted in the AMANDA-B10 analysis.

\section{SEARCH FOR CASCADES WITH AMANDA-B10 and -II}

We have performed a full-reconstruction search for the Cherenkov light
patterns resulting from electromagnetic or hadronic showers ({\it
cascades}) induced by a diffuse flux of high-energy extraterrestrial
neutrinos.  Data collected by AMANDA-B10 in 1997 was used.
Demonstrating $\nu$-induced cascade sensitivity is an important step
for neutrino astronomy because the cascade channel probes all neutrino
flavors, whereas the muon channel is primarily sensitive to $\nu_\mu$.
Compared to muons, cascades provide more accurate energy measurement
and better separation from background, but they suffer from worse
angular resolution and reduced effective volume.  In addition, it is
straightforward to calibrate the cascade response of neutrino
telescopes such as AMANDA through use of, for example, {\it in-situ}
light sources.  As with muons, cascades become increasingly easier to
identify and reconstruct as detector volumes get larger.

The electron neutrino produces cascades via the charged current
interaction and all neutrino flavors produce cascades via the neutral
current interaction. Cascade-like events are also produced in charged
current $\nu_\tau$ interactions.

After application of simple selection criteria to reduce the data
sample size while preserving any potential signal events, cascade
vertex position, time of production, energy and direction are
reconstructed using several maximum likelihood functions that take
into account the Cherenkov emission, absorption and scattering of
light~\cite{picrc,kowalski:diplm,taboada:phd,amanda-cascades}.  This
full-reconstruction approach is less susceptible to spurious
backgrounds than other techniques.

In the absence of a tagged source of high energy neutrino-induced
cascades, we rely on {\it in-situ} light sources, catastrophic energy
losses by down-going cosmic ray muons, and Monte Carlo simulations to
understand the response of the detector.  The successful
reconstruction of pulsed laser data and the reconstruction of isolated
catastrophic muon energy losses, described in~\cite{amanda-cascades},
demonstrate that the detector is sensitive to high energy cascades.

The 90\% C.L. limit on the diffuse flux of $\nu_e+\nu_\mu+\nu_\tau+
\overline{\nu}_e+\overline{\nu}_\mu+\overline{\nu}_\tau$ for
neutrino energies between 5~TeV and 300~TeV, assuming a neutrino flux
ratio of 1:1:1 at the detector, is:

\begin{equation}
   \label{nu_l_limit_eq}
   E^2\frac{d\Phi}{dE} < 9.8 \times 10^{-6} \; \mathrm{GeV\,cm^{-2}\,s^{-1}\,sr^{-1}}.
\end{equation}

\noindent The 90\% C.L. limit on the diffuse flux of $\nu_e+\overline{\nu}_e$ for
neutrino energies between 5~TeV and 300~TeV
is:
\begin{equation}
   \label{nu_e_limit_eq}
   E^2\frac{d\Phi}{dE} < 6.5 \times 10^{-6} \;
   \mathrm{GeV\,cm^{-2}\,s^{-1}\,sr^{-1}}. 
\end{equation}
The latter limit is independent of the assumed neutrino flux ratio.
(Note
that since the limit in Eq.~\ref{nu_l_limit_eq} is on the sum of the
fluxes of all neutrino flavors, and the limit in
Eq.~\ref{nu_e_limit_eq} is on an individual flavor, the former limit
should be divided by a factor of three to compare it properly to the
latter.)

\begin{figure}
\includegraphics[width=0.46\textwidth]{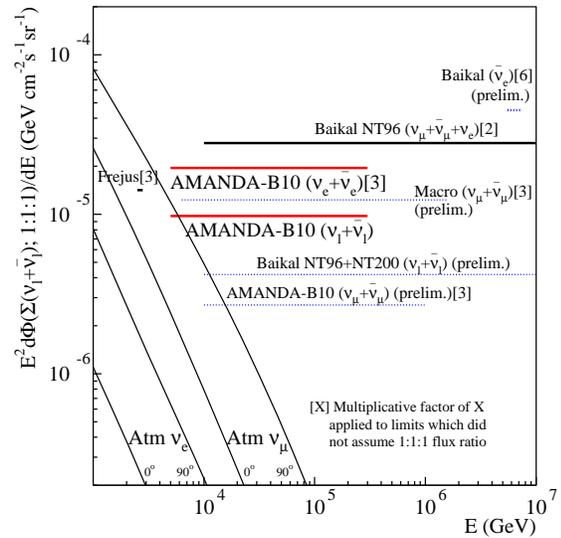}
\caption{\label{fig:limits} 
The limits on the cascade-producing neutrino flux, summed over the
three active flavors, presented in this work and in other experiments,
with multiplicative factors applied as indicated to permit comparison
of limits derived with different assumed neutrino fluxes at the
detector: Baikal ($\overline{\nu}_e$)~\cite{venice01:baikal} (at the
$W^\pm$ resonance); Baikal NT96
($\nu_\mu+\overline{\nu}_\mu+\nu_e$)~\cite{baikal:astropart}; Frejus
($\nu_\mu+\overline{\nu}_\mu$)~\cite{frejus}; MACRO
($\nu_\mu+\overline{\nu}_\mu$)~\cite{icrc01:macro}.  Baikal NT96+NT200
($\nu_l+\overline{\nu}_l$)~\cite{venice01:baikal,jan:baikal}; AMANDA B-10
($\nu_\mu+\overline{\nu}_\mu$)~\cite{icrc01:dif}; Also shown are the
predicted horizontal and vertical $\nu_e$ and $\nu_\mu$ atmospheric
fluxes~\cite{lipari}.}
\end{figure}

Our results together with other limits on the flux of diffuse
neutrinos are shown in fig.~\ref{fig:limits}.  Since recent results
from other low energy neutrino
experiments~\cite{sno-cc,sno-nc,sno-dn,superk-atm} indicate that
high-energy extragalactic neutrinos will have a neutrino flavor flux
ratio of 1:1:1 upon detection, in this figure we scale limits derived
under different assumptions accordingly.  For example, to do a
side-by-side comparison of a limit on the flux of
$\nu_e+\nu_\mu+\nu_\tau+\overline{\nu}_e+\overline{\nu}_\mu+\overline{\nu}_\tau$,
derived under the assumption of a ratio of 1:1:1, to a limit on just
the flux of $\nu_\mu+\overline{\nu}_\mu$, the latter must be degraded
by a factor of three.  (N.B.: We assume that
$\nu$:$\overline{\nu}$::1:1, and we take into account the different cross sections
for $\nu$ and $\overline{\nu}$.)

It should be noted that most searches of diffuse fluxes shown in
fig.~\ref{fig:limits} are based on the observation of up-going
neutrino-induced muons. Only Baikal and AMANDA have presented limits
from analyses that search for neutrino-induced cascades and only the
AMANDA analysis uses full cascade event reconstruction.

Data acquired by AMANDA-II is currently under study and, as with the
analysis of atmospheric neutrinos, preliminary results from that work
clearly demonstrate the enhanced power of the larger AMANDA-II
detector.  Angular acceptance improves to nearly $4\pi$, backgrounds
are much easier to reject, and energy acceptance improves by a factor
of three to $E_\nu \sim 1$~PeV.  In accordance with our blind analysis procedures,
we presently use just 20\% of the AMANDA-II data from the year 2000,
and obtain a preliminary limit which is lower than that described
above by roughly a factor of 2--3.  For a UHE neutrino flux at the
current best limit~\cite{icrc01:dif}, we would expect to detect about
eight UHE-neutrino-induced cascade events in the full 2000 dataset, on
an expected background of less than one event.  This analysis is
described in more detail in a contributed poster at this
conference~\cite{mareks-poster}.

\section{SEARCH FOR UHE $\nu_\mu$ FROM POINT SOURCES WITH AMANDA-II}

We have conducted a general search for continuous emission of muon
neutrinos from a spatially localized direction in the northern sky.
Backgrounds are reduced by requiring a statistically significant
enhancement in the number of reconstructed upward-going muons in a
small bin in solid angle.  Furthermore, the background for a
particular bin can be calculated from the data by averaging over the
data external to that bin in the same declination band.  In
contrast to other searches, this search is more tolerant of the
presence of background, so the signal is optimized on $S/\sqrt{B}$,
where $S$ represents the signal and $B$ the background, rather than on
$S/B$, which emphasizes signal purity.

Data acquired by AMANDA-B10 in 1997 has been analyzed and the results
presented in~\cite{b10-point-source}.  With AMANDA-II data taken in
2000, we gain improved sensitivity to events near the horizon since
the detector has double the number of PMTs and a larger lever arm in
the horizontal dimension.  Assuming a customary $E^{-2}$ power law
spectrum at the source, and a flux of
$10^{-8}$~GeV$\,$cm$^{-2}$s$^{-1}$sr$^{-1}$, we predict approximately two
signal and one background events in a $6^\circ \times 6^\circ$ angular
bin.  Preliminary sensitivities to various point sources are given in
Table~\ref{table:point_source_sensitivities}.  In order to achieve
blindness in this analysis the right ascension of each event (i.e., its azimuthal
angle) has been scrambled (at the South Pole this effectively
scrambles the event time), and they will only be unscrambled once all
selection criterion have been set.
\begin{table*}[htb]
\caption{Preliminary estimated sensitivities of AMANDA-II to various 
         point sources in data taken in the year 2000.  The sensitivity
         is defined as the predicted average limit from an
         ensemble of experiments with no signal, and is calculated using
         background levels predicted from off-source data.}
\label{table:point_source_sensitivities}
\renewcommand{\tabcolsep}{2pc} % enlarge column spacing
\renewcommand{\arraystretch}{1.2} % enlarge line spacing
\begin{tabular}{llll} \hline
Source              & \multicolumn{1}{c}{Declination} 
                            & \multicolumn{1}{c}{$\mu$ ($\times 10^{-15}$cm$^{-2}$s$^{-1}$)} 
                                   & \multicolumn{1}{c}{$\nu$ ($\times 10^{-8}$cm$^{-2}$s$^{-1}$)}\\ \hline
%%%Markarian 421       &       & 2.6  & 1.1 \\
SS433               & 5.0   & 11.0 & 3.6 \\
Crab                & 22.0  & 4.0  & 1.9 \\
Markarian 501       & 39.8  & 2.5  & 1.5 \\
Cygnus X-3          & 41.5  & 2.6  & 1.6 \\
Cass. A             & 58.8  & 2.1  & 1.5 \\  \hline
\end{tabular}\\[2pt]
\end{table*}

\section{SEARCH FOR $\nu_\mu$ FROM DIFFUSE SOURCES WITH AMANDA-II}

The search for diffuse sources of UHE $\nu_\mu$--induced muons is
similar to the analysis used to detect atmospheric $\nu_\mu$--induced
muons, as both analyses require a sample of events with low contamination from
misreconstructed downward-going atmospheric muons.
Since high-energy muons will deposit more energy in the
detector volume than low-energy muons, the diffuse analysis further
requires that events have a high channel density, $\rho_{\rm ch} > 3$,
where the channel density is defined as the number of hit channels per
10~m tracklength.  The background in the signal region is estimated
by extrapolating from lower-energy data satisfying $\rho_{\rm ch} < 3$.

\begin{figure}[htb]
\includegraphics*[scale=0.42,bb=30 30 560 560]{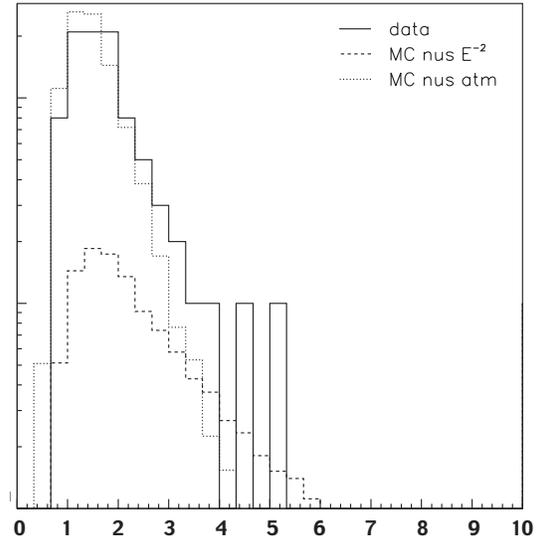}
\caption{Distributions of the channel density $\rho_{\rm ch}$ for data, 
         simulated signal and simulated background.  The simulated signal
         assumes a customary $E^{-2}$ power law spectrum at the source
         and a neutrino flux of
         $10^{-6}\,\mathrm{GeV\,cm^{-2}\,s^{-1}\,sr^{-1}}$.  Events are
         kept if they satisfy $\rho_{\rm ch} > 3$.}
\label{fig:aii-diffuse}
\end{figure}

Using a 20\% subsample of the AMANDA-II data from 2000, we detect 6
events satisfying all selection criteria.  Simulations indicate that
we would detect 3.0 events from a UHE neutrino flux at the current
best limit~\cite{icrc01:dif}, assuming a customary $E^{-2}$ power law
spectrum at the source, and 1.9 events from atmospheric neutrino
interactions.  (N.B. We use a subsample of the data in order to
achieve blindness in this analysis.)  The distributions of $\rho_{\rm
ch}$ for data, simulated signal and simulated background are shown in
fig.~\ref{fig:aii-diffuse}.

The predicted average limit from an ensemble of experiments with no
signal, or {\it sensitivity}, is roughly $1.3 \times
10^{-6}$~GeV$\,$cm$^{-2}$s$^{-1}$sr$^{-1}$, and the preliminary limit is less
than roughly $10^{-6}$~GeV$\,$cm$^{-2}$s$^{-1}$sr$^{-1}$.  This is about the
same as the limit obtained with the {\it full} sample of AMANDA-B10
data from 1997.

\section{SEARCH FOR $\nu_\mu$ FROM GRBs WITH AMANDA-B10 and -II}

The search for UHE $\nu_\mu$--induced muons from gamma-ray bursts
(GRBs) leverages temporal and directional information from
satellite-based observations of GRB photons to realize a nearly
background-free analysis.  Assuming the predicted
spectrum~\cite{grb-spectrum}, we search for muon neutrinos in the
energy range of 10~TeV--100~PeV and use off-time data to
estimate background and to achieve blindness in the analysis.  

\begin{figure}[htb]
\vspace{9pt}
% \framebox[55mm]{\rule[-21mm]{0mm}{43mm}}
\includegraphics[scale=0.45]{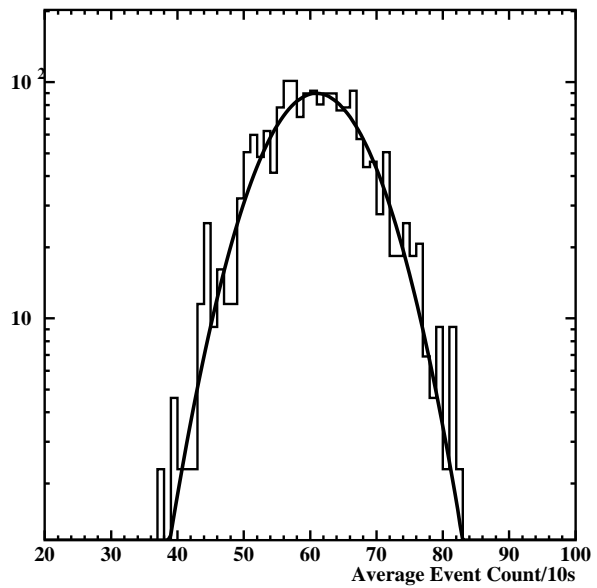}
\caption{The average event count per 10~s period, demonstrating the high stability of the AMANDA-II data.
         Some basic event selection criteria have been applied.}
\label{fig:aii-stability}
\end{figure}

The analysis looks for enhancements in the rate of upward-going muons
in time windows coinciding with the reported GRB ``T90'' time
window~\cite{t-ninety}, which can range from roughly 2~s to 50~s in
width, and in the reported GRB direction.  Detector stability over
these time scales is therefore an important measure of how effective
this analysis can be.  Figure~\ref{fig:aii-stability} shows the count
rate per 10~s bin in a time window of roughly $\pm 1$~hour around a
particular GRB, using events which have passed certain basic selection
criteria.  The good agreement with a Gaussian distribution shows
that the detector did not experience instrumental effects which in
principle could mimic a GRB.  Plots for all the other GRB in the
sample exhibit the same Gaussian behavior.  In the AMANDA data
spanning the years 1997--2000, we anticipate having a sample of
roughly 500 GRBs to search for neutrino emission.

\section{CONCLUSIONS}

The Antarctic Muon and Neutrino Detector Array is a functioning
neutrino telescope which has been used to search for a variety of
interesting possible sources of ultrahigh energy neutrino flux.  Data
from AMANDA-B10 have been searched not only for upward-going muons but
also for electromagnetic and hadronic showers using full event
reconstruction for the first time.  Data from AMANDA-II have been
searched in preliminary analyses for $\nu_\mu$-induced muons and
$\nu_x$-induced showers, and in all instances the enhanced sensitivity
of the enlarged detector is clearly evident.  We therefore have
confidence that AMANDA-II analyses will yield significantly improved
limits over those published using AMANDA-B10 data--or perhaps the
analyses will yield the discovery of UHE neutrinos.

\section{ACKNOWLEDGMENTS}

This research was supported by the following agencies: U.S.  National
Science Foundation, Office of Polar Programs; U.S. National Science
Foundation, Physics Division; University of Wisconsin Alumni Research
Foundation; U.S. Department of Energy; Swedish Research Council;
Swedish Polar Research Secretariat; Knut and Alice Wallenberg
Foundation, Sweden; German Ministry for Education and Research; U.S.
National Energy Research Scientific Computing Center (supported by the
Office of Energy Research of the U.S.  Department of Energy);
UC-Irvine AENEAS Supercomputer Facility; Deutsche
Forschungsgemeinschaft (DFG); Flanders Institute to Encourage
Scientific and Technological Research in Industry (FNRS-FWO, IWT);
Belgian Federal Office for Scientific, Technical and Cultural affairs
(OSTC), Belgium. D.F.C. acknowledges the support of the NSF CAREER
program.


\begin{thebibliography}{9}

   \bibitem{uhe-nu-hep-theories} See, for example: V. Bertin, E. Negri
   and J. Orloff, submitted to European Phy. J. C (2002)
   hep-ph/0204135; C.T. Hill, D.N. Schramm and T.P. Walker,
   Phys. Rev. {\bf D36} (1987) 1007; P. Bhattacharjee, C.T. Hill and
   D.N. Schramm, Phys. Rev. Lett. 69 (1992) 567.

   \bibitem{uhe-nu-astro-theories} See, for example: T.K. Gaisser,
   F. Halzer and T. Stanev, Phys. Rept. 258 (1995) 173; E. Waxman and
   J. Bahcall, Phys. Rev. {\bf D59} (1999) 023002; K. Mannheim,
   R.J. Protheroe and J.P. Rachen, Phys. Rev. {\bf D63} (2000) 023003;
   J.G. Learned and K. Mannheim, Ann. Rev. Nucl. Part. Sci. 50 (2000)
   679; V.S. Berezinsky and V.I. Dokuchaev, Astropart. Phys. 15 (2001)
   87; A. Levinson and E. Waxman, Phys. Rev. Lett. 87 (2001)
   1711101.01.

   \bibitem{other-experiments} G. Domogatsky for the Baikal
   Collaboration and J. Carr, contributed talks, this conference.

   \bibitem{ice-properties} P.B. Price, K. Woschnagg, and D. Chirkin, Geophys. Res. Lett. 27 (2000) 2129.
   \bibitem{albrechts-contribution} A. Karle for the IceCube Collaboration, contributed talk, this conference.
   \bibitem{nature} E. Andres {\it et al.},  Nature {\bf 410}, 6827, 441-443, 2001.
   \bibitem{b10-atmnu} J. Ahrens {\it et al.}, Phys. Rev. D (2002) 012005.
   \bibitem{b10-wimps} J. Ahrens {\it et al.}, Phys. Rev. D (2002) 032006.
   \bibitem{b10-monopoles} P. Niessen and C. Spiering for the AMANDA collaboration, in Proc. of ICRC, 
            Hamburg, Germany, 2001, p. 1496.
   \bibitem{b10-ehe} S. Hundertmark for the AMANDA Collaboration, contributed poster I-1, this conference;
           S. Hundertmark for the AMANDA Collaboration,  in 
           Proc. Second Workshop Methodical Aspects Underwater/Underice Neutrino Telescopes, Hamburg, 2001.
   \bibitem{b10-point-source} J. Ahrens {\it et al.}, Submitted to Astrophysical Journal, 2002.
   \bibitem{b10-diffuse} G.C.~Hill and M.J.~Leuthold for the AMANDA Collaboration, in 
           Proc. of ICRC, Hamburg, Germany, 2001, p. 1113.
   \bibitem{b10-supernova} J. Ahrens {\it et al.}, Astropart. Phys. {\bf 16}, 345-359, 2001.
   \bibitem{picrc} M. Kowalski and I. Taboada for the AMANDA Collaboration, in Proc. Second Workshop Methodical Aspects
           Underwater/Ice Neutrino Telescopes, Hamburg, Germany, 2001.
   \bibitem{kowalski:diplm} M. Kowalski, Diploma Thesis, Humboldt University, Berlin, Germany, 2000.
   \bibitem{taboada:phd} I. Taboada, Ph.D. Thesis, University of Pennsylvania, 2002.
   \bibitem{amanda-cascades} J. Ahrens {\it et al.}, Submitted to Phys. Rev. D, 2002.
   \bibitem{sno-cc} Q.R. Ahmad {\it et al.}, Phys. Rev. Lett. 87 (2001) 071301.
   \bibitem{sno-nc} Q.R. Ahmad {\it et al.}, Phys. Rev. Lett. 89 (2002) 011301.
   \bibitem{sno-dn} Q.R. Ahmad {\it et al.}, Phys. Rev. Lett. 89 (2002) 011302.
   \bibitem{superk-atm} Fukuda {\it et al.}, Phys. Rev. Lett. 85 (2000)  3999.
   \bibitem{venice01:baikal} V. Balkanov {\it et al.}, in Proc. 9th Int. Workshop on Neutrino Telescopes, 
           Venice, Italy, 2001, Vol. 3 p. 591.
   \bibitem{baikal:astropart} V. Balkanov {\it et al.}, Astro. Part. Phys. (2000) vol. 14, p. 61.
   \bibitem{frejus} W.~Rhode  {\it et al.}, Astro. Part. Phys. (1994) vol. 4, p. 217.
   \bibitem{icrc01:macro} Perrone {\it et al.}, in Proc. 27th Int. Cosmic Ray Conf., Hamburg, Germany, 2001, p. 1073.
   \bibitem{jan:baikal} J. Dzhilkibaev, private communication, 2002.
   \bibitem{icrc01:dif} G. Hill and M. Leuthold for the AMANDA collaboration, 
           in Proc. 27th Int. Cosmic Ray Conf., Hamburg, Germany, 2001, p. 1113.
   \bibitem{lipari} P. Lipari, Astro. Part. Phys. 1 (1993) 193.
   \bibitem{mareks-poster} M. Kowalski for the AMANDA collaboration, contributed poster I-2, this conference.
   \bibitem{grb-spectrum} E. Waxman and J. Bahcall, Phys. Rev. Lett. 78 (1997) 2292.
   \bibitem{t-ninety} W.S. Paciesas {\it et al.}, Ap. J. Suppl. (1999) vol. 122,  p. 465. 

\end{thebibliography}
\end{document}